\documentclass[fleqn]{mnras}

\usepackage{newtxtext,newtxmath}
\usepackage{multirow} 
\usepackage{tikz}
\usepackage{pgfplots}
\usepackage{adjustbox} 

\usepackage[T1]{fontenc}

\DeclareRobustCommand{\VAN}[3]{#2}
\let\VANthebibliography\thebibliography
\def\thebibliography{\DeclareRobustCommand{\VAN}[3]{##3}\VANthebibliography}

\usepackage{graphicx}	
\usepackage{amsmath}	

\title[EEGMobile]{Sleep Stage Classification Using a Pre-trained Deep Learning Model}

\author[Hassan Ardeshir, Mohammad Araghi]{
Hassan Ardeshir,$^{1}$
Mohammad Araghi$^{2}$
\\
$^{1}$Department of Computer Science, University of Tehran, Tehran, Iran\\
$^{2}$Department of Computer Engineering, University of Tehran, Tehran, Iran
}

\pubyear{2023}
\usepackage[square,sort,comma,numbers]{natbib}

\makeatletter
\renewcommand{\@biblabel}[1]{\hfill\relax[#1]}
\renewcommand\@cite[1]{\textsuperscript{[#1]}}
\makeatother

\pgfplotsset{compat=1.18}

\begin{document}
\label{firstpage}
\pagerange{\pageref{firstpage}--\pageref{lastpage}}
\maketitle

\begin{abstract}
One of the common human diseases is sleep disorders. The classification of sleep stages plays a fundamental role in diagnosing sleep disorders, monitoring treatment effectiveness, and understanding the relationship between sleep stages and various health conditions. A precise and efficient classification of these stages can significantly enhance our understanding of sleep-related phenomena and ultimately lead to improved health outcomes and disease treatment.

Models others propose are often time-consuming and lack sufficient accuracy, especially in stage N1. The main objective of this research is to present a machine-learning model called "EEGMobile". This model utilizes pre-trained models and learns from electroencephalogram (EEG) spectrograms of brain signals. The model achieved an accuracy of 86.97\% on a publicly available dataset named "Sleep-EDF20", outperforming other models proposed by different researchers. Moreover, it recorded an accuracy of 56.4\% in stage N1, which is better than other models. These findings demonstrate that this model has the potential to achieve better results for the treatment of this disease.
\end{abstract}




\section{Introduction}

\subsection{Sleep}
Sleep is a reversible state in which eyes are closed, and most parts of the body are inactive, allowing the individual to become unconscious and providing an opportunity for the body to recover energy and alleviate fatigue and anxiety. Sleep is considered a fundamental human function, comprising about one-third of our lives.

Research indicates that sleep is vital for strengthening learning and memory, as the brain forms and reinforces new learning pathways during sleep. Additionally, adequate sleep enhances problem-solving abilities and improves creativity. Dreams during sleep also play a significant role in memory consolidation and brain processing
\cite{1}.

The behavior and decisions of individuals during the day are dependent on the duration and quality of their sleep. Chronic sleep deprivation can lead to various mental problems, such as cognitive disorders, stress, and depression, significantly impacting a person's life. Moreover, insufficient sleep can have broad consequences on physical health, including increased risks of obesity, diabetes, high blood pressure, cancer, and cardiovascular diseases
\cite{2}.

Sleep consists of distinct stages that individuals go through during a night's rest. These stages are characterized by specific patterns of brain activity, eye movements, and muscle activity. The two main categories of sleep stages are Rapid Eye Movement (REM) and Non-Rapid Eye Movement (NREM), with NREM further divided into N1, N2, and N3 stages.

In general, each human sleep includes multiple sleep cycles, each containing these stages. Some sleep cycles may not include all the stages, and, for example, a cycle may only consist of N1 and N2 stages
\cite{3}.

\subsection{EEG}
Brain signals, also known as neural signals or brainwaves, are electrical activities generated by brain neurons. These signals are produced due to the communication between different regions of the brain. Measuring brain signals is crucial for understanding brain function, and cognition, and investigating various neurological conditions. The brain, being the controller of the body and particularly the nervous system, is the most essential part of the human body, and studying it can aid in understanding and treating various physical and mental illnesses.

The presence of electrical currents and waves in the brain was discovered in 1875 by an English physician, Richard Caton, and over the past century, significant advancements have been made in studying these brainwaves. One of the primary techniques used to measure brain signals is EEG Electroencephalography (EEG). EEG is a non-invasive medical imaging technique that records the brain's electrical activity from the scalp using metal electrodes and conductive gels
\cite{4}.

\subsection{EEG \& Sleep Stages}
As mentioned earlier, sleep disorder treatment centers identify the root cause of sleep problems by examining brain function and initiating appropriate measures for resolution. One of the methods used to assess brain function is EEG. Doctors record an individual's brain activity using EEG during a sleep cycle and analyze the stages and cycles of sleep. Based on the differences observed compared to the normal state, they provide their diagnosis and prescribe medications or necessary interventions for patients.

Therefore, the most crucial aspect of treating sleep disorders lies in accurately diagnosing the sleep stages during an EEG test.

\subsection{Spectrograms}
Spectrography has significant applications in sound analysis. In essence, an audio signal is represented as a waveform that indicates changes in amplitude over time. However, a spectrogram illustrates the changes in frequency of the waveform over time, with amplitude represented as the third dimension using color. Thus, the vertical axis represents frequency in Hertz, and the horizontal axis represents time.

All spectrograms are not created equal; an algorithm called Fast Fourier Transform (FFT) is commonly used to compute these spectrograms. In FFT, a parameter called size (or the number of data points involved) is variable, leading to different outcomes. Generally, higher FFT sizes provide finer frequency details, known as frequency resolution, while lower FFT sizes provide finer time details, known as temporal resolution.

For instance, if you want to identify microphone noise, a higher FFT size is helpful. On the other hand, if you want to detect a high-frequency event, you should opt for a smaller FFT size. Hence, spectrograms can be used to eliminate noise or unwanted sounds.

Indeed, as mentioned earlier, the output of a spectrogram is a color image. In recent years, models working on audio, particularly systems converting speech to text, first transform the audio signal into its spectrogram representation. They then operate on the spectrogram using methods like CNNs or pre-trained models that excel at image processing \cite{5}.

\section{Transfer Learning}
\subsection{Pre-trained models}
Image processing plays a vital role within the broader scope of image analysis and computer vision systems. The outcomes of image processing exert significant influence over subsequent high-level tasks, facilitating the recognition and understanding of image data. Deep learning has emerged as a potent tool for tackling low-level vision tasks, such as image super-resolution, inpainting, deraining, and colorization, in recent times. While these image-processing tasks share commonalities, there has been limited exploration of pretraining models across these domains.

The application of pretraining holds promise for addressing two key challenges in image processing. First, task-specific datasets can be constrained, especially in scenarios involving sensitive or costly data. For instance, this is evident in medical image processing, which includes tasks ranging from the segmentation of specific anatomical structures and the detection of lesions to the differentiation between pathological and healthy tissue in various organs \cite{10}. Another example is the use of satellite images for assessing the quality of resulting images in urban areas \cite{11}.

Additionally, factors like varying camera parameters, lighting conditions, and weather can introduce significant variations in training data distributions. Second, the specific image processing tasks required are often unknown until the test image is presented. Consequently, a suite of image processing modules must be prepared, each with distinct objectives but potential for shared underlying operations.

The concept of pretraining has already gained traction in natural language processing and computer vision.
For instance, many object detection models utilize pre-trained backbones from ImageNet classification \cite{14}.

A wealth of well-established networks can be found online, with AlexNet setting the foundation for convolutional neural networks \cite{16}, VGGNet known for its uniform depth \cite{17}, and ResNet's groundbreaking skip connections that have transformed deep learning, notably improving computer vision tasks \cite{18}.

 In the realm of natural language processing, Transformer-based models have revolutionized tasks like translation and question-answering. Their success hinges on pretraining these models on vast text corpora, followed by fine-tuning them on task-specific datasets.

Efforts have been made to extend the triumph of Transformers into the domain of computer vision, marking an exciting intersection of these two fields.

\subsection{Medical Application}
Medical imaging plays an important role in the medical
area and is a powerful tool for diagnosis. With the
development of computer technology such as machine
learning, computer-aided diagnosis has become a popular
and promising direction. Note that medical images are
generated by special medical equipment, and their labeling
often relies on experienced doctors. Therefore, in many
cases, it is expensive and hard to collect sufficient training
data. Transfer learning technology can be utilized for medical imaging analysis. A commonly used transfer learning
approach is to pre-train a neural network on the source
domain (e.g., 
 ImageNet is a large-scale ontology of images built upon the backbone of the WordNet structure, encompassing a total of 3.2 million images. It is significantly more accurate than current image datasets which is useful for three simple applications: object recognition, image classification, and automatic object clustering \cite{9}) and then fine-tuning it
based on the instances from the target domain.

\subsection{MobileNetV3}
MobileNetV3 encompasses two distinct models, MobileNetV3-Large and MobileNetV3-Small, designed to cater to high and low resource usage scenarios, respectively. These models are the result of platform-aware Neural Architecture Search (NAS) and NetAdapt techniques, which refine and optimize network architectures for improved performance \cite{24}.

In the realm of computer vision, recent advancements have introduced convolutional neural network architectures that prioritize both speed and size efficiency. These pivotal computer vision architectures, including NASNet \cite{25}, MobileNets \cite{26, 27}, EfficientNet \cite{28}, MnasNet \cite{29}, and ShuffleNets \cite{30}, are acclaimed for their swift training processes. NASNet automates architecture search for tailored image solutions, while MobileNets prioritize mobility and efficiency, excelling in real-time image analysis. EfficientNet balances size and depth for versatile image applications, MnasNet combines architecture search with mobile-friendly design, and ShuffleNets reduce computation overhead, ideal for real-time video processing and edge computing, collectively advancing diverse computer vision domains \cite{31, 34}.

These networks implement depthwise convolutions, a technique where convolutional kernels are applied individually to each input channel, enhancing the extraction of spatial information. Notably, these depthwise convolutional kernels are shared across all input channels, thus optimizing model efficiency and lowering computational costs. It's important to note that learning the size of these depthwise convolutional kernels can pose challenges, potentially increasing the intricacy of training processes.

One of the noteworthy recent contributions in this domain is the MobileNetV3 architecture, which has demonstrated significant advancements in computer vision tasks \cite{35}.

\begin{figure}
    \centering
    \includegraphics[width=\columnwidth]{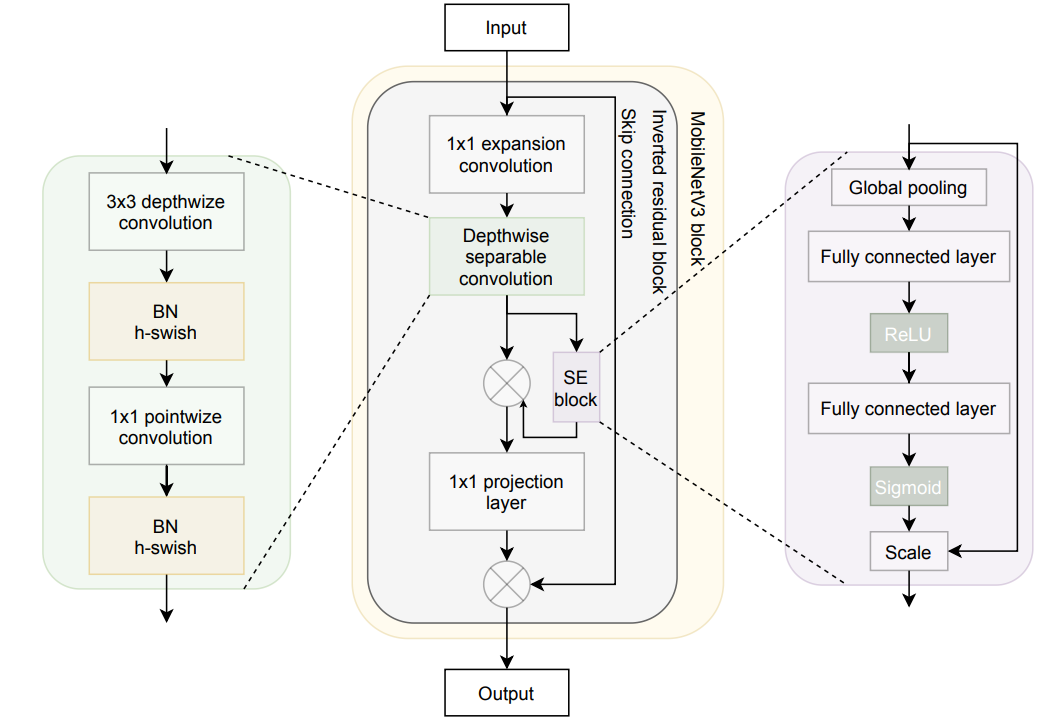}
    \caption{The structure of MobileNetV3 blocks and components \protect\cite{38}.}
    \label{fig:mobilenetv3}
\end{figure}

MobileNetV3, an evolution of MobileNetV1 and MobileNetV2, represents a significant leap in mobile-friendly neural network architectures. Howard et al. introduced this version using Network Architecture Search (NAS) with the NetAdapt algorithm, aiming to optimize MobileNet for low-resource hardware platforms in terms of size, performance, and latency. The architecture enhancements in MobileNetV3, depicted in Figure \ref{fig:mobilenetv3}, draw inspiration from its predecessors.

A notable addition to MobileNetV3 is the introduction of a novel nonlinearity known as "hard swish" (h-swish), a modified variant of the sigmoid function from a previous work \cite{36}. The h-swish non-linearity 
 minimizes training parameters, thereby reducing model complexity and size.

Within the MobileNetV3 block, a core component emerges—the inverted residual block. This block combines a depthwise separable convolution block with a squeeze-and-excitation block \cite{29}, drawing inspiration from bottleneck blocks \cite{37}. The inverted residual connection links input and output features within the same channels, enhancing feature representations while conserving memory.

The depth-wise separable convolutional operation comprises a depthwise convolutional kernel applied to each channel, followed by a 1 × 1 pointwise convolutional kernel with batch normalization (BN) and ReLU or h-swish activation functions. This transformation replaces the traditional convolutional block and reduces model capacity.

To further optimize training, MobileNetV3 incorporates a squeeze-and-excitation (SE) block, which selectively emphasizes relevant features in each channel \cite{38}. 

\section{Materials and Methods}
\subsection{Dataset}
In this section, we will discuss the examination of the sleep-edf dataset, which is relevant to this topic, as well as the proposed method and other methods provided by others, which are learned through it.

This dataset is accessible through the PhysioNet website, and the article related to how this dataset was collected has also been reviewed.

This dataset consists of 197 recorded files of Polysomnographic (PSG) data related to the entire night's sleep of both healthy and diseased individuals (those with sleep disorders). Each PSG includes EEG signals in two channels, Pz-Oz and Fpz-Cz, as well as EOG signals from individuals' brains, with a recording frequency of 100 Hz. Additionally, the sleep stage is determined every 30 seconds. The dataset is labeled with "ST" for patients with sleep disorders and "SC" for healthy individuals. It should be noted that a maximum of two files have been recorded for each individual, corresponding to two different nights.

In this dataset, an alternative naming convention for sleep stages has been used, consisting of a cycle of Wake, S1, S2, S3, S4, and REM. Essentially, a 6-stage cycle is considered, and in this new classification, S3 and S4 correspond to N3. Therefore, in most other works, including the method presented in this report, the naming convention is initially changed as follows:
\begin{equation}
S1 \rightarrow N1, \quad S2 \rightarrow N2, \quad S3, S4 \rightarrow N3
\end{equation}
In works that have been done to solve this problem, three subsets of this dataset are considered:
\begin{description}
\item[$\ast$ \textbf{Sleep-EDF8:}]
This subset, published by the PhysioNet website in 2013, includes only 8 recorded files related to 4 healthy individuals and 4 patients. As a result, the number of data samples (30-second segments) is 15188.

\item[$\ast$ \textbf{Sleep-EDF20:}]
In this subset, 39 files related to 20 healthy individuals (initial 20 healthy individuals of the original dataset) are selected, resulting in a total of 42308 data samples.

\item[$\ast$ \textbf{Sleep-EDF78:}]
In this subset, all 153 files related to 78 healthy individuals (all healthy individuals from the original dataset) are considered, resulting in a total of 195479 data samples.
\end{description}

It is evident that in all subsets, the number of samples in stages is not balanced with each other, which can affect the training outcome, causing the designed model to predict the stage with more samples more frequently.

\begin{figure}
    \centering
    \includegraphics[width=\columnwidth]{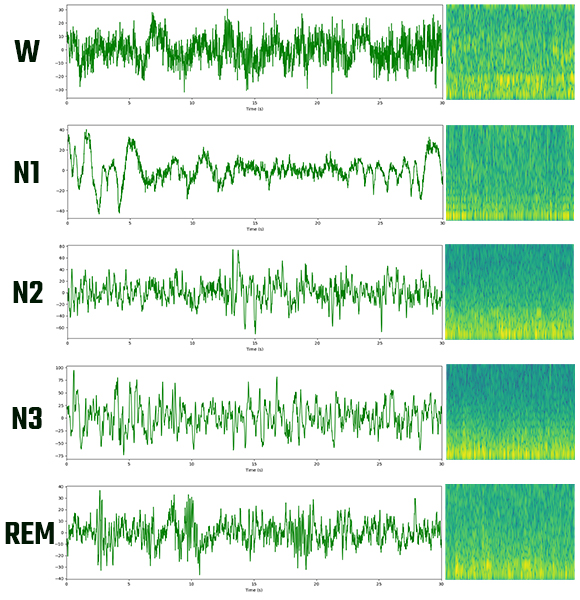}
    \caption{EEG signal with a frequency of 100 Hz and corresponding spectrograms}
    \label{fig:eegspec}
\end{figure}

Spectrograms can reflect the activity of a specific frequency. When the EEG signal is stronger, the color of its corresponding spectrogram is brighter (yellow), and the weaker the signal is, the darker the color (blue) \cite{6}. 

With the improved visualization of frequency in Figure \ref{fig:eegspec}, we can now enhance our ability to define sleep stages using non-linear measures:
\begin{description}
\item[$\ast$ \textbf{Wake:}]
During the wakeful stage, you are fully conscious and alert. Your eyes are open, and your brain activity, as measured by EEG, is characterized by rapid fluctuations. In this stage, there is prominent beta activity, which is characterized by a frequency range of 13-26 Hz and a low voltage of 10-30 $\mu$V. Additionally, there may be some alpha activity, ranging from 8-12 Hz, with a higher voltage of 20-40 $\mu$V. This is the stage where you are actively engaged with the world around you, and your thoughts are typically alert and clear.

\item[$\ast$ \textbf{N1 (Drowsiness):}]
As you transition from wakefulness to sleep, you enter the drowsiness stage. During this phase, your eye movements may slow down, and you may experience slow movements of eye-rolling. Notably, the alpha waves, which were present during wakefulness, start to disappear. In their place, theta waves emerge, typically in the range of 4-7 Hz. These theta waves are indicative of a transitional state between wakefulness and light sleep. You may start to feel more relaxed, and your thoughts might become less coherent as you drift towards sleep.

\item[$\ast$ \textbf{N2 (Light Speed):}]
In the stage of light sleep, your eye movements cease, and you become more detached from your surroundings. This stage is characterized by distinctive patterns on the EEG. Bursts of brain activity are visible, and you may see the emergence of sleep spindles, which are short bursts of oscillatory brain waves in the range of 11-15 Hz, as well as K-complexes. These features are superimposed on a background of theta waves. Light sleep serves as a transitional phase, where your body and mind begin to relax further in preparation for deeper sleep.

\item[$\ast$ \textbf{N3 (Deep Sleep):}]
Deep sleep is a crucial phase for physical and mental restoration. Delta waves make their appearance slowly in this stage, and they are characterized by a low frequency of 1-3 Hz and a high EEG amplitude exceeding 75 $\mu$V. Sleep spindles and K-complexes may still be present but are less prominent compared to earlier stages. Deep sleep is vital for physical healing, immune function, and memory consolidation. It is often challenging to wake someone from a deep sleep, and if you are awakened, you may feel disoriented initially.

\item[$\ast$ \textbf{REM Sleep:}]
REM sleep is a unique stage characterized by rapid eye movements, as the name suggests. During REM sleep, your muscles are temporarily paralyzed to prevent you from acting out your dreams. The EEG activity during REM sleep is mixed in frequency, and it is often associated with low voltage. Occasional bursts of sawtooth waves may appear on the EEG. This stage is where most vivid dreaming occurs, and brain activity resembles wakefulness in some ways, despite being associated with muscle atonia. REM sleep is vital for emotional processing, memory consolidation, and overall cognitive function \cite{15}.
\end{description}

\subsection{Method}
\subsubsection{Overview}
In alignment with the study conducted by \cite{6}, our focus centers exclusively on the \textit{Fpz-Cz} channel within the \textit{EEG} signals derived from the \textit{Sleep-EDF} dataset. These samples transform spectrograms, following which we construct a novel model utilizing the pre-existing \textit{MobileNetV3Large} architecture. Subsequently, we perform fine-tuning on specific layers of this model, resulting in the creation of \textit{EEGMobile}. This newly devised model is then trained to employ the generated spectrograms. Given the prior training of the \textit{MobileNetV3Large} component on images and the inherent resemblance between spectrograms and images, we have attained promising outcomes within this domain, which will be subjected to a comprehensive analysis in subsequent sections.

\subsubsection{Preprocess}
We store the spectrograms of the data samples under examination, and to calculate these spectrograms, we employ the \texttt{scipy} library in Python.

Specifically, the Spectrogram function (Figure \ref{fig:CodeSpect}) within the aforementioned library is configured with five input parameters, as documented in \cite{7}:

\begin{figure}
	\centering
	\includegraphics[width=\columnwidth]{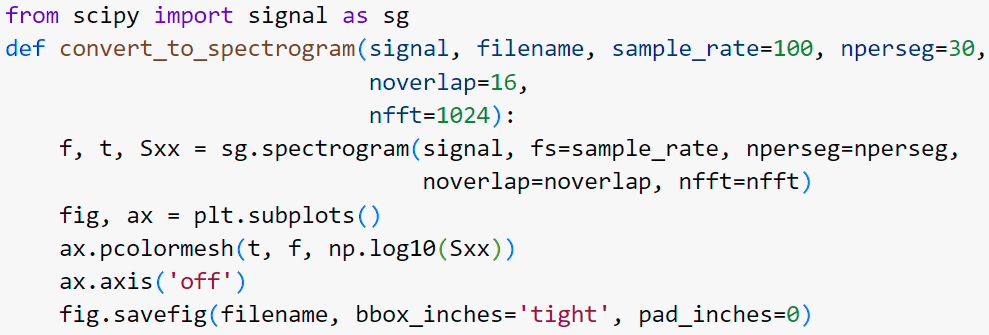}
	\caption{Converting a signal to its corresponding image}
    \label{fig:CodeSpect}
\end{figure}

\begin{description}
\item[\b{signal}]
Input signal represented as a time series with measured values.
\item[$\ast$ \textbf{fs}]
Recording frequency of the input time series, which, as mentioned earlier, should be set to 100 for our dataset.
\item[$\ast$ \textbf{nperseg}]
The length of each input signal should be set to 30 as each signal is 30 seconds long.
\item[$\ast$ \textbf{noverlap}]
It is the Number of points that have overlap between segments. In the spectrogram section, this is discussed with an example, and it is set to 16. However, changing this value might yield better results as it significantly affects the visual characteristics.
\item[$\ast$ \textbf{nfft}]
Length of the FFT used, set to 1024. It could have been lower, but in our case, a smaller value didn't yield good results for the model we used. The resulting image didn't resemble an image but merely separate colored points.
\end{description}

It is noteworthy that, in contrast to the referenced study \cite{6}, there is no requirement to crop the generated spectrogram images. The code automatically eliminates graph-related elements when saving the image, resulting in an output image with dimensions of 497 pixels in length and 369 pixels in width.

Furthermore, when we take into account the scale of processing 195,479 images and acknowledge that this process is I/O-bound, the read-and-write operations to the hard drive can substantially impede program performance. To mitigate this issue, we leverage Python's ThreadPoolExecutor to concurrently handle these tasks for multiple images, resulting in an eight-fold acceleration compared to the non-concurrent approach.

To eliminate the need for generating spectrograms from signals during subsequent experiments and to have consistent inputs, these images are stored.

Additionally, while reading the images again, since we plan to use 20-fold cross-validation, we group samples related to a single subject into one fold. For example, in the \texttt{Sleep-EDF8} dataset, we ultimately have 20 folds, each containing a total of 52609 data points.

Furthermore, because the constructed network performs better on images of size 224 * 224 when reading the images, they are resized using the \texttt{resize} command from the \texttt{cv2} library in Python \cite{8}.

\subsubsection{EEGMobile}
To construct this artificial network, the Keras library has been employed. Now we will provide a detailed description of the constructed network.

In this network, we begin with the \textit{MobileNetV3Large}, a pre-trained artificial network on images, which we introduced earlier. The output of this network is a 4-dimensional tensor.

Following that, we have the fine-tuning section, which allows us to adjust the output provided by the pre-trained network to suit our problem. This section requires further training and modifications. In this section, an initial \textit{Global Average Pooling2D} layer is included to transform the mentioned 4-dimensional output to 2-dimensional while reducing the number of dimensions.

Subsequently, a \textit{Dense} layer with dimensions 224 and a \textit{relu} activation function is placed, serving to consolidate previous information.

Further down, a \textit{Dense} layer with 5 dimensions and a \textit{softmax} activation function is present. These dimensions correspond to the classes present in the dataset: \textit{(Awake, N1, N2, N3, REM)}. The \textit{softmax} function ensures that the output represents the class with the highest probability.

Now, we can configure our neural network in a way that some of its layers won't undergo training. This reduces the number of trainable parameters and consequently increases the training speed for each iteration. For instance, we can make the initial layers of the pre-trained model untrainable. This significantly speeds up the training process as a whole. However, according to the experiments conducted, which we will elaborate on later, it's better to allow these layers to be trained at least for a few epochs initially.

Consequently, we create the aforementioned network using the \textit{adam} optimizer and the \textit{sparse categorical crossentropy} loss function, which is suitable for models with multiple classes. Subsequently, we can proceed to train this network.

\section{Training Experiments}
\subsection{Details}
In this experiment, only the \textit{Sleep-EDF20} dataset is considered. Therefore, as in other related articles on this topic, we utilize a \textit{20-fold cross-validation}. In each training stage, the data of one individual is designated as the validation set. A new model is constructed and trained using data from the remaining 19 individuals. The model's performance is then evaluated with the validation dataset, repeating this process for all 20 individuals. Ultimately, the results are averaged to provide the final output.

It's important to mention that the TensorFlow library is employed to process data concurrently with a graphics card, significantly enhancing the speed of model training.

As a result, with a batch size of 16 and only 20 epochs, we were able to effectively train our neural network. However, it's worth noting that the initial 5 epochs involved training the entire network, including all layers of both the pre-trained model and the added fine-tuning layers. In the subsequent 15 epochs, only the last layers (the final 15 layers of the pre-trained model and the fine-tuning layers) were trained. This approach greatly accelerated the training process, and within a very short time, using a \textit{GTX 1060 6GB} graphics card, we achieved the desired outcome. This result was expected, as in recent years, pre-trained models have consistently outperformed self-constructed models.

Also, We tested three different methods to read data to train the model:
\begin{description}
    \item[$\ast$ \textbf{HDD (Hard Disk Drive)}]
Spectrograms are stored on the HDD with sufficient storage capacity but slower read/write speeds. The Keras module processes images on the GPU, which accelerates training. However, each batch loads data into RAM, introducing latency compared to faster storage.
    \item[$\ast$ \textbf{SSD (Solid-State Drive):}]
We use SSDs for improved speed, as they offer significantly faster read/write speeds than HDDs. SSDs are a good choice for GPU-based training, though they are more expensive. They are cost-effective for datasets like sleep-edf8.
    \item[$\ast$ \textbf{RAM (Random Access Memory):}]
Storing the entire dataset in RAM provides the fastest training speeds. RAM's speed surpasses HDDs and SSDs. Importantly, it eliminates the need to read from slower storage during each epoch, speeding up training. RAM doesn't require a capacity equivalent to the dataset size. Moreover, initial data loading into RAM can utilize multi-threading for even faster data access.

\end{description}
\subsection{Result}
The model achieved an average accuracy of 86.97\%. The comparison of our model's performance with other experiments is presented in Table \ref{table:13}.

The proposed model performs even better than EEGSNet \cite{6}, especially in the N1 class. It's worth mentioning that the number of trainable parameters is set to 3.2 million, assuming all layers are trainable. However, when only the final fifteen layers of the pre-trained model, along with the fine-tuning layers, are considered trainable, the count reduces to 525,829 which is less than EEGSNet's parameters. Also, EEGSNet's experiments were performed on a server with four NVIDIA Tesla V100-DGXS GPUs, considering the best performance after 3 times. However, we ran our experiment using a GTX 1060 6GB graphics card only one time.

\begin{table}
	\centering
 \Huge
    \renewcommand{\arraystretch}{1.8}
	\caption{Final Results of \textit{EEGMobile} on \textit{Sleep-EDF20} Dataset}
	\begin{adjustbox}{max width=\columnwidth}
		\begin{tabular}{|c|c|c|c|c|c|c|c|c|c|c|}
			\hline
			\multirow{2}{*}{Methods} & \multirow{2}{*}{Param} & \multirow{2}{*}{Epochs} & \multicolumn{3}{c|}{Overall Metrics (\%)} & \multicolumn{5}{c|}{Per-Class F1 (\%)} \\
			& & & ACC & MF1 & Kappa & W & N1 & N2 & N3 & REM \\
			\hline
			DeepSleepNet \cite{39} & 24.7 M & 41950 & 82 & 76.9 & 0.76 & 84.7 & 46.6 & 85.9 & 84.8 & 82.4 \\
			TinySleepNet \cite{40} & 1.3 M & 42308 & 85.4 & 80.5 & 0.80 & 90.1 & 51.4 & 88.5 & 88.3 & 84.3 \\
			EEGSNet \cite{6} & 0.6 M & 42308 & 86.82 & 81.57 & 0.82 & 90.76 & 52.41 & 88.78 & 87.0 & 87.89 \\
			\hline
			EEGMobile & 3.2 M & 20 & 86.97 & 81.42 & 0.81 & 95.4 & 56.4 & 88.86 & 82.82 & 83.64 \\
			\hline
		\end{tabular}
	\end{adjustbox}
 	\label{table:13}
\end{table}

\begin{figure}
	\centering
	\includegraphics[width=\columnwidth]{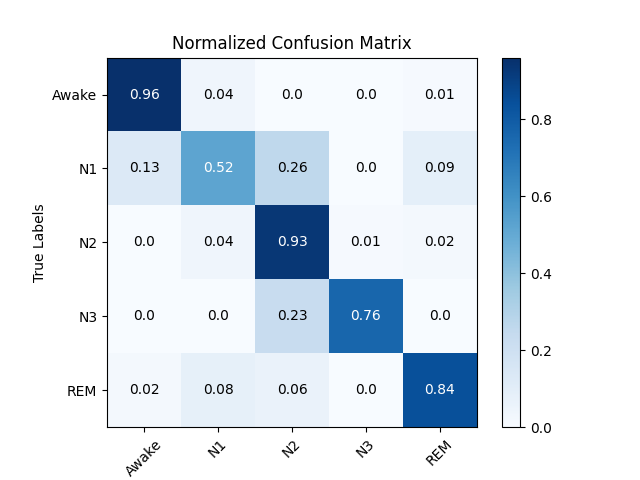}
	\caption{Confusion Matrix of the Proposed Method (\textit{EEGMobile})}
\end{figure}

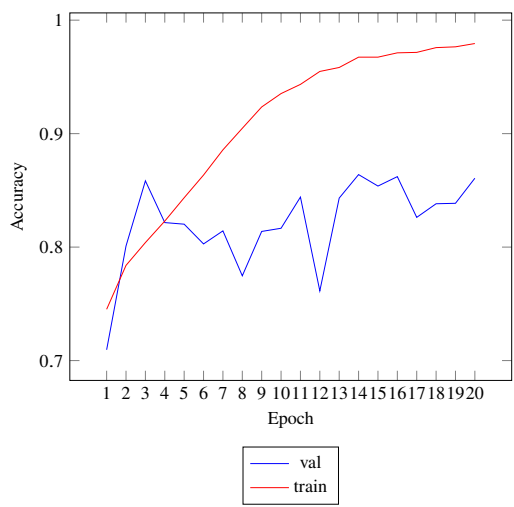
\begin{figure}
    \hspace{8pt}
    \begin{adjustbox}{max width=190pt}
    \begin{tikzpicture}
        \begin{axis}[
            xlabel={Epoch},
            ylabel={Accuracy},
            legend style={at={(0.5,-0.18)}, anchor=north},
            xtick={1,2,3,4,5,6,7,8,9,10,11,12,13,14,15,16,17,18,19,20},
            ]
            
            \addplot[mark=none,blue] coordinates {
                (1, 0.7095)
                (2, 0.8010)
                (3, 0.8584)
                (4, 0.8216)
                (5, 0.8202)
                (6, 0.8028)
                (7, 0.8143)
                (8, 0.7747)
                (9, 0.8138)
                (10, 0.8166)
                (11, 0.8441)
                (12, 0.7614)
                (13, 0.8432)
                (14, 0.8639)
                (15, 0.8538)
                (16, 0.8621)
                (17, 0.8262)
                (18, 0.8382)
                (19, 0.8386)
                (20, 0.8607)
            };
            
            \addplot[mark=none,red] coordinates {
                (1, 0.7452)
                (2, 0.7839)
                (3, 0.8040)
                (4, 0.8229)
                (5, 0.8435)
                (6, 0.8635)
                (7, 0.8857)
                (8, 0.9047)
                (9, 0.9235)
                (10, 0.9353)
                (11, 0.9434)
                (12, 0.9548)
                (13, 0.9583)
                (14, 0.9674)
                (15, 0.9674)
                (16, 0.9712)
                (17, 0.9716)
                (18, 0.9758)
                (19, 0.9765)
                (20, 0.9794)
            };
            \legend{val, train}
        \end{axis}
    \end{tikzpicture}
    \end{adjustbox}

    \caption{Validation vs. Train Accuracy}
\end{figure}

Additionally, we can observe the varying results obtained through different reading methods:
\begin{description}
    \item[$\ast$ \textbf{HDD:}]
On HDD, each epoch took 9 minutes, and with 20 epochs, one fold was completed in 180 minutes. The entire project yielded results after 60 hours.
    \item[$\ast$ \textbf{SSD:}]
With SSD, each epoch took 4 minutes, and with 20 epochs, one fold was completed in 80 minutes. The entire project yielded results after 26 hours.
    \item[$\ast$ \textbf{RAM:}]
In contrast, RAM only required 5 minutes to store all data initially. Subsequently, each epoch was completed in just 1 minute. Thus, each fold was done in 20 minutes, and the entire training process concluded in 400 minutes.
\end{description}
In summary, RAM significantly enhances the training speed, evident from the swift completion of each epoch and fold. Furthermore, the speed benefits of RAM become even more apparent with faster RAM, allowing for the possibility of training the model with additional epochs. With the availability of greater RAM capacity, the project becomes scalable, making it feasible to tackle larger datasets efficiently.

\section{Conclusions}
The results of the study and experimentation on the proposed model have shown promising performance with a higher accuracy rate compared to the works of others in the Sleep-EDF20 dataset. It has also exhibited significantly better speed. Nevertheless, for further validation and generalization of our findings, we recommend extending the evaluation to datasets such as Sleep-EDF8, Sleep-EDF78, and SHHS. This will allow us to compare the experimental results on different datasets with other works.

Furthermore, to uncover the full potential of the model, conducting experiments with an increased number of training iterations (similar to other works with around 1000 iterations) is suggested. Undoubtedly, increasing the number of iterations can make the model more accurate and help it converge to a better optimum.

In conclusion, in this report, we conducted a comprehensive study on sleep stage classification using the Fpz-Cz channel in EEG signals. Our goal was to design a robust and efficient model capable of accurately classifying different sleep stages using the American Academy of Sleep Medicine classification system. We presented a model that learns, in which a pre-trained model is fine-tuned and augmented with carefully selected layers. The proposed model has shown promising results in sleep stage classification tasks.




\bibliographystyle{plain}

\label{lastpage}
\end{document}